\documentclass[aps,twocolumn,superscriptaddress,groupedaddress]{revtex4-1}  

\include{packages.tex}
\usepackage{graphicx}  
\usepackage{dcolumn}   
\usepackage{bm}        
\usepackage{amssymb}   
\usepackage[utf8]{inputenc}  
\usepackage[T1]{fontenc} 
\usepackage{amsmath}

\begin{document}

\title{On the origin of the giant spin detection efficiency \\ in tunnel barrier based electrical spin detector}

\author{E. Fourneau}
\affiliation    {Solid-State Physics – Interfaces and Nanostructures, Q-MAT, CESAM, University of Liège,  Liège, 4000, Belgium}

\author{A. V. Silhanek}
\affiliation    {Experimental Physics of Nanostructured Materials, Q-MAT, CESAM, University of Liège, Liège, 4000, Belgium}

\author{N. D. Nguyen}
\affiliation    {Solid-State Physics – Interfaces and Nanostructures, Q-MAT, CESAM, University of Liège,  Liège, 4000, Belgium}

\date{\today}

\begin{abstract}
Efficient conversion of a spin signal into an electric voltage in mainstream semiconductors is one of the grand challenges of spintronics. This process is commonly achieved via a ferromagnetic tunnel barrier where non-linear electric transport occurs. In this work, we demonstrate that non-linearity may lead to a spin-to-charge conversion efficiency larger than 10 times the spin polarization of the tunnel barrier when the latter is under bias of a few mV. We identify the underlying mechanisms responsible for this remarkably efficient spin detection as the tunnel barrier deformation and the conduction band shift resulting from a change of applied voltage. In addition, we derive an approximate analytical expression for the detector spin sensitivity $P_{\textrm{\tiny det}}(V)$. Calculations performed for different barrier shapes show that this enhancement is present in oxide barriers as well as in Schottky tunnel barriers even if the dominant mechanisms differs with the barrier type. Moreover, although the spin signal is reduced at high temperatures, it remains superior to the value predicted by the linear model. Our findings shed light into the interpretation and understanding of electrical spin detection experiments and open new paths to optimize the performance of spin transport devices. 
\end{abstract}
\maketitle

\section{Introduction}

Injection, transport and detection of a spin-polarized current in a non-magnetic (NM) semiconductor (SC) are cornerstones of spintronics. In the past years, encouraging results were obtained on spin-polarized current injection into mainstream group-IV semiconductors such as Si \cite{Spiesser2017,Dankert2013}, Ge \cite{Spiesser2014,Zhou2011} or SiC \cite{Huang2018} at room temperature, as well as into other promising materials such as graphene \cite{Ohishi2007}. To that purpose, ferromagnetic (FM) tunnel junctions are widely regarded as one of the best approaches to both generate a spin polarization into a SC and convert it into a voltage signal. This is in part due to the limited spin absorption in the FM and the reduced resistance mismatch \cite{Fert2001}. Nowadays, the literature is abondant on works quantifying the performance of a FM tunnel barrier to generate and detect a spin signal in a SC as well as in evaluation of the spin lifetime \cite{Jansen2012}. Promising results have been reported over the last decade with a surprising and unexpected spin detection efficiency, demonstrating in some cases a pick up voltage higher than the injected spin signal \cite{Spiesser2014,Widmann2014}. This outstanding spin detection was reported for devices where the FM tunnel contact used for detection was biased, as in technologically relevant devices \cite{Datta1990,Datta2018,Fujita2019,Spiesser2019,Oltscher2017,sato2017}. This amplification effect offers an interesting perspective for on-chip integration of spin-based circuits and has attracted considerable attention from theoretical standpoint. Indeed, various tentative mechanisms were proposed over the recent years to explain the observed large spin detection efficiency, such as two-step tunneling \cite{Tran2009}, thermionic emission \cite{Jansen2015} or lateral current inhomogeneity \cite{Dash2009}, to name just a few. Unfortunately, none of the above mentioned mechanisms seems to satisfactorily account for all the experimental findings.

Recently, it has been experimentally highlighted that the discrepancies between experiments and theoretical calculations are due to the energy dependence of the carrier transmission probability in the tunnel junction. This suggests that a new description based on a non-linear transport of spin should be invoked \cite{Jansen2018}. Although preliminary ideas in this direction were already advanced some ten years ago, it is not until recently that non-linearities have been recognized as an essential ingredient for the understanding of FM tunnel junctions. 
In this work, we aim at identifying the implications of non-linearity in FM tunnel junctions using a theoretical approach, including crucial aspects overlooked in previous studies while directly responsible for the giant spin detection efficiency. In this way, we are able to explain how a spin signal may be converted into a charge signal with an efficiency of several hundred of percent. It is worth noting that the obtained results apply to any spin detection devices composed of a FM/SC contact with a tunnel barrier in between, thus including not only oxide and tunnel barriers, but also various pseudo-substrates such as graphene and other 2D materials.

\section{Linear Model of spin detection}
\begin{figure*}
\includegraphics[width=0.99\textwidth]{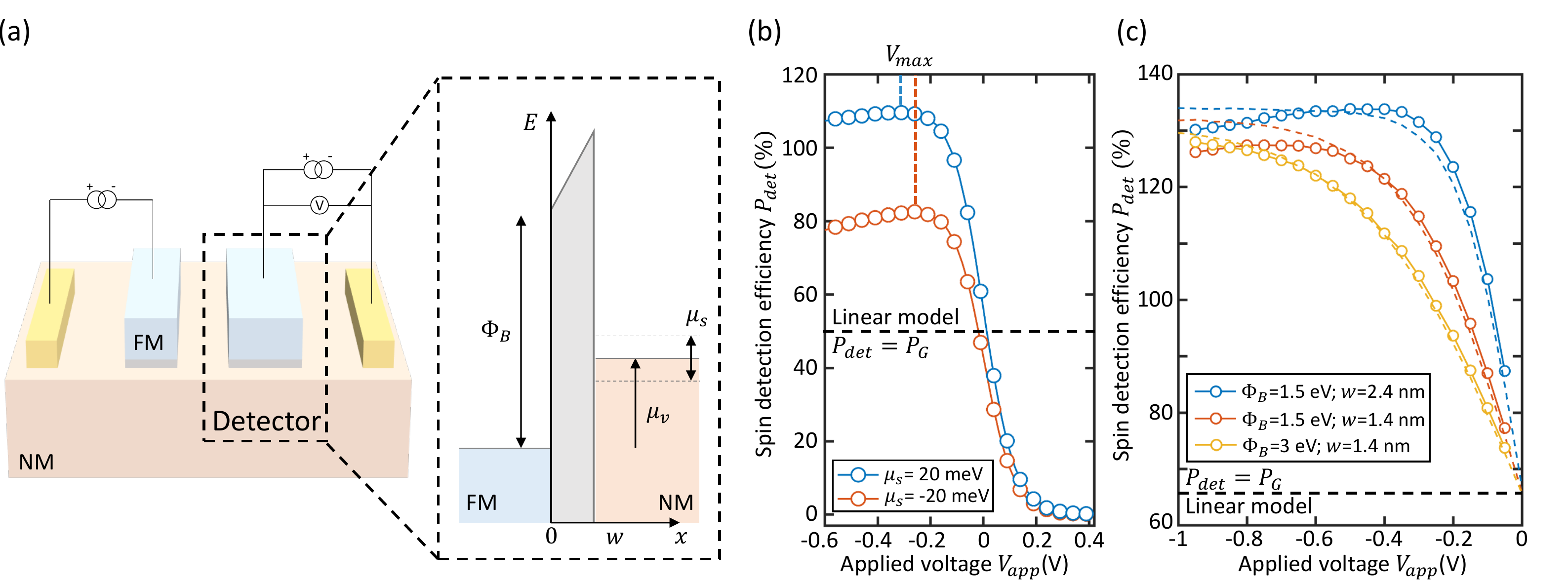}
\caption{Nonlinear spin detection efficiency under bias. (a) Schematic energy band diagram of the FM/B/NM tunnel contact under a bias, where $\mu_v=-eV_{\textrm{\tiny app}}$ is the applied electrochemical potential, $w$ and $\Phi_B$ are the width and the height of the barrier. A drawing of the spintronic device is also provided for clarity. (b) Computed spin detection efficiency with tunnel bias for two different spin accumulations $\mu_s$ ($w = 4$ nm; $\Phi_B=1$ eV; $P_G=50\%$; $T=1$ K). The dashed line is obtained by cancelling the barrier deformation under bias, corresponding to the linear model. (c) Spin detection efficiency as a function of applied voltage for different barriers. Dashed lines correspond to results reported in Ref. \cite{Jansen2018}.}
\label{modelFMBNM}
\end{figure*}

The measurement of a voltage variation due to the presence of a spin polarization is called electrical spin detection. This mechanism is achieved by the transfer of carriers through a spin-dependent barrier (B) formed at the interface between a ferromagnetic (FM) layer and  a non-magnetic (NM) layer. In this section, we briefly introduce the model based on a linear transport description through a tunnel barrier sandwiched between a FM and a NM metal \cite{Fert2001,Jansen2012}. The model assumes that the conductance of the barrier depends on the spin orientation. We denote by $G_+$ and $G_-$ the conductance, respectively for spin parallel and anti-parallel to the main magnetization of the FM contact. The application of a voltage bias $V_{\textrm{\tiny app}}\equiv-\mu_v/e$, $\mu_v$ being the NM electrochemical potential and $e$ the elementary charge, will force the transfer of charge carriers from one side of the interface to the other side with a spin preference. In this model, it is assumed that a preferential spin population is generated in the NM via an external mechanism, leading to  a splitting of the electrochemical potential $\mu_s =\mu_+ - \mu_-$ at the B/NM interface (see Fig. \ref{modelFMBNM}(a)). In presence of spin accumulation, the spin-dependent current density through the barrier, i. e. from the NM into the FM, is given by
\begin{equation}
J_\pm = \frac{G_\pm}{A}\left( \frac{\mu_v}{e}\pm \frac{\mu_s}{2e} \right),
\end{equation}
where $A$ is the junction cross sectional area. The total current is 
\begin{equation}
J=J_+ + J_- =\frac{G}{Ae}\left(\mu_v +P_G\frac{\mu_s}{2}\right),
\end{equation}
with $G=G_++G_-$ the total conductance barrier and $P_G=\frac{G_+-G_-}{G}$ the spin polarization of the tunnel conductance. The tunnel junction bias is
\begin{equation}
V_{\textrm{\tiny app}}=-\frac{\mu_v}{e}=-\frac{JA}{G}+P_G\frac{\mu_s}{2e}
\label{eqJ}
\end{equation}
The last term of Eq. (\ref{eqJ}) represents a voltage signal due to the spin accumulation and is called spin voltage $V_{\textrm{\tiny spin}}=-P_G\mu_s/(2e)$. The figure of merit of a device designed for spin detection, namely spin detection efficiency, is defined as the electrical voltage generated at the detector per unit of spin accumulation, $P_{\textrm{\tiny det}}=-V_{\textrm{\tiny spin}}/(\mu_s/2e)$ which equals $P_G$ in the linear model. 

Since the linear model for spin transport is based on the Ohm's law, the tunnel barrier is therefore considered as a conductance $G$ independent of any applied voltage. This consideration is unrealistic as tunnelling process is strongly non-linear and is the essence of the spin filtering effect. 

\section{Non-linear Model}

Although the linear theory for spin injection and detection captures the essential mechanisms of these processes, deviations from the linear model have been systematically reported in experiments based on 3-terminal (3T) Hanle devices and in 4T devices where both injector and detector are under bias \cite{Jansen2012,Fujita2019}. Recently, Jansen \textit{et al}. \cite{Jansen2018} experimentally observed that the spin detection efficiency $P_{\textrm{\tiny det}}$ at a tunnel junction strongly depends on the applied bias, offering a way to magnify the detected spin accumulation. In their work, the authors pointed out that the spin signal amplification is inherent to the non-linear transport occurring at the tunnel junction which arises from the dependence of the transmission probability with the energy of injected carriers. 
In simple words, since the transmission probability increases with energy, the preferential spin population will undergo a higher impact than the minority spin population when changing the applied voltage. Naturally, the increase of $P_{\textrm{\tiny det}}$ with the applied voltage could be attributed mainly to this effect.   


While this reasoning seems to qualitatively capture the trend observed in most experimental results, there still exists some features that remain unexplained \cite{Jansen2018}. Notably in their explication of the origin of the non-linearity, the fact that the increase of applied voltage needed to compensate the loss of current after spin precession (i. e., the spin voltage) becomes sensitive to the increase of junction bias, leading to a $P_{\textrm{\tiny det}}$ \textit{independent} on the bias. Moreover, their model looks limited as it may not be able to justify why a signal differing of several order of magnitude is pointed out in many spin tunnel devices \cite{Jansen2012}. In this work, we aim at investigating the mechanisms governing this non-linear transport by using an incremental approach, which in complexity is progressively added to the model. 
We focus first on the simplest model of a tunnel barrier between two metals (FM/B/NM) where the effects of the barrier deformation under bias and the importance of the spin accumulation intensity are thoughtfully investigated. Then, the NM metals is replaced by a highly degenerate SC in order to evaluate the impact of a band gap, the degeneracy level and the energy dependence of the density of states (DOS). At a later stage, we address the variation of the tunnel barrier shape to compare the detection efficiency of oxide and Schottky tunnel barriers. 

Calculations were performed by solving the spin-dependent non-linear tunnel transport equations based on the two-channel model, as described in Refs. \cite{VanSon1987,Fert2001}. In a first approximation, a simplified version of the free-electron description is used for a semi-classical approach \cite{Wolf1989}:
\begin{equation}
\begin{split}
&J_\pm =\\
&\frac{4\pi m_e^2}{h^3}\frac{1\pm P_G}{2}\int_{-\infty}^\infty T(E)\left[f_{\pm,sc}(E,V_{\textrm{\tiny app}},\mu_s)-f_{fm}(E) \right]dE
\end{split}
\end{equation}
where $h$ is the Planck's constant and $m_e$ is the effective electron mass. Using the WKB approximation, the transmission function is given by  \cite{Razeghi2009}
\begin{equation}
T(E,V_{\textrm{\tiny app}})=\exp\left( -\frac{4m_ee}{h^2}\int_0^w\sqrt{ \phi(x,V_{\textrm{\tiny app}})-E }~dx\right) 
\end{equation}
while the Fermi-Dirac distributions $f_+$ and $f_-$ are the spin-dependent functions
\begin{equation}
f_{\pm,sc}(E,V_{\textrm{\tiny app}},\mu_s)=\left[1+\exp\left(\dfrac{E + eV_{\textrm{\tiny app}}\pm\mu_s/2}{k_B T}\right)\right]^{-1},
\end{equation}
with $k_B$ the Boltzman's constant and $T$ the absolute temperature. It is assumed that the spin dependence of the carrier transport has two origins. Firstly, the DOS in the FM layer is spin-dependent. In our calculations, we hypothesize that this difference does not vary with the energy (free electron model) and corresponds to the spin polarity $P_G$ in order to fit with the linear model. Secondly, the density of carriers with spin up and down in the NM layer will differ even if the DOS is spin-independent (also energy-independent). This effect is due to the presence of a spin accumulation and is expressed in the Fermi-Dirac distribution.
The spin voltage is obtained by equalizing the total current under a given bias $V_{\textrm{\tiny app}}$ in the presence of a spin accumulation $\mu_s$ in the NM layer, with the total current corresponding to a bias $V_{\textrm{\tiny app}}+V_{\textrm{\tiny spin}}$ in absence of spin accumulation:
\begin{equation}
J\left( V_{\textrm{\tiny app}},\mu_s \right) = J\left( V_{\textrm{\tiny app}}+V_{\textrm{\tiny spin}},0 \right)
\label{eqJspin}
\end{equation}

\subsection{Origin of the non-linear dependence $P_{\textrm{\tiny det}}(V_{app})$}

We first investigate the case of a rectangular tunnel barrier (B) in a three-layer stack FM/B/NM (Fig. \ref{modelFMBNM}(a)). This model has already been studied previously \cite{Jansen2018}, however, in the present work the FM quasi Fermi level is set as the reference electrode and the oxide barrier deformation with applied voltage is considered.

As illustrated in Fig. \ref{modelFMBNM}(b), the spin detection efficiency $P_{\textrm{\tiny det}}$ strongly depends on the junction bias. At 0 V, the spin detection bears the value that is predicted by the linear model. However, when the structure is under a non-zero external bias, $P_{\textrm{\tiny det}}$ strongly deviates from the linear model, increasing or decreasing in magnitude for negative bias (spin extraction regime) and positive bias (spin injection regime), respectively. While the the general behaviour of $P_{\textrm{\tiny det}}$ as function of $V_{\textrm{\tiny app}}$ is similar to the results of Jansen \textit{et al.} \cite{Jansen2018}, our findings provide further unanticipated features. First of all, the non-linear effect does not show a perfect odd symmetry and $P_{\textrm{\tiny det}}(V)$ exhibits a non-monotonic dependence (not due to a change of $P_G$), leading to a maximal detection efficiency for a specific voltage $V_{\textrm{\tiny max}}$ . Secondly, the maximal value of $P_{\textrm{\tiny det}}$ is not limited to $2 P_G$. Moreover, the value depends on the tunnel barrier dimensions  as it will be demonstrated below. Finally, $P_{\textrm{\tiny det}}$ is very sensitive to the intensity of the spin accumulation in the vicinity of the barrier.

Concerning the origin of this non-linearity, it has been previously suggested that the general behaviour of $P_{\textrm{\tiny det}}$ as function of $V_{\textrm{\tiny app}}$ for both injection and extraction regimes may be related to the steepness of the energy dependency of the barrier transmission function $dT(E)/dE$ \cite{Jansen2018}. Indeed, the spin accumulation mainly affects the transport of electrons of higher energy. As the bias decreases ($V_{\textrm{\tiny app}}<0$), high energy electrons have an increasingly dominating contribution to the device current as consequence of the exponential energy dependence of $T(E)$, thus leading to a higher impact of the spin accumulation on the carrier transport. On the other hand, for positive biases, the spin accumulation impacts the tail of the transmission function, resulting in a weaker perturbation of the transport through the barrier. Consequently, the more proeminent is the slope of $T(E)$, the more important is the non-linearity of the spin detection.

\subsection{Effect of the barrier deformation}

\begin{figure}[h!]
\includegraphics[width=0.49\textwidth]{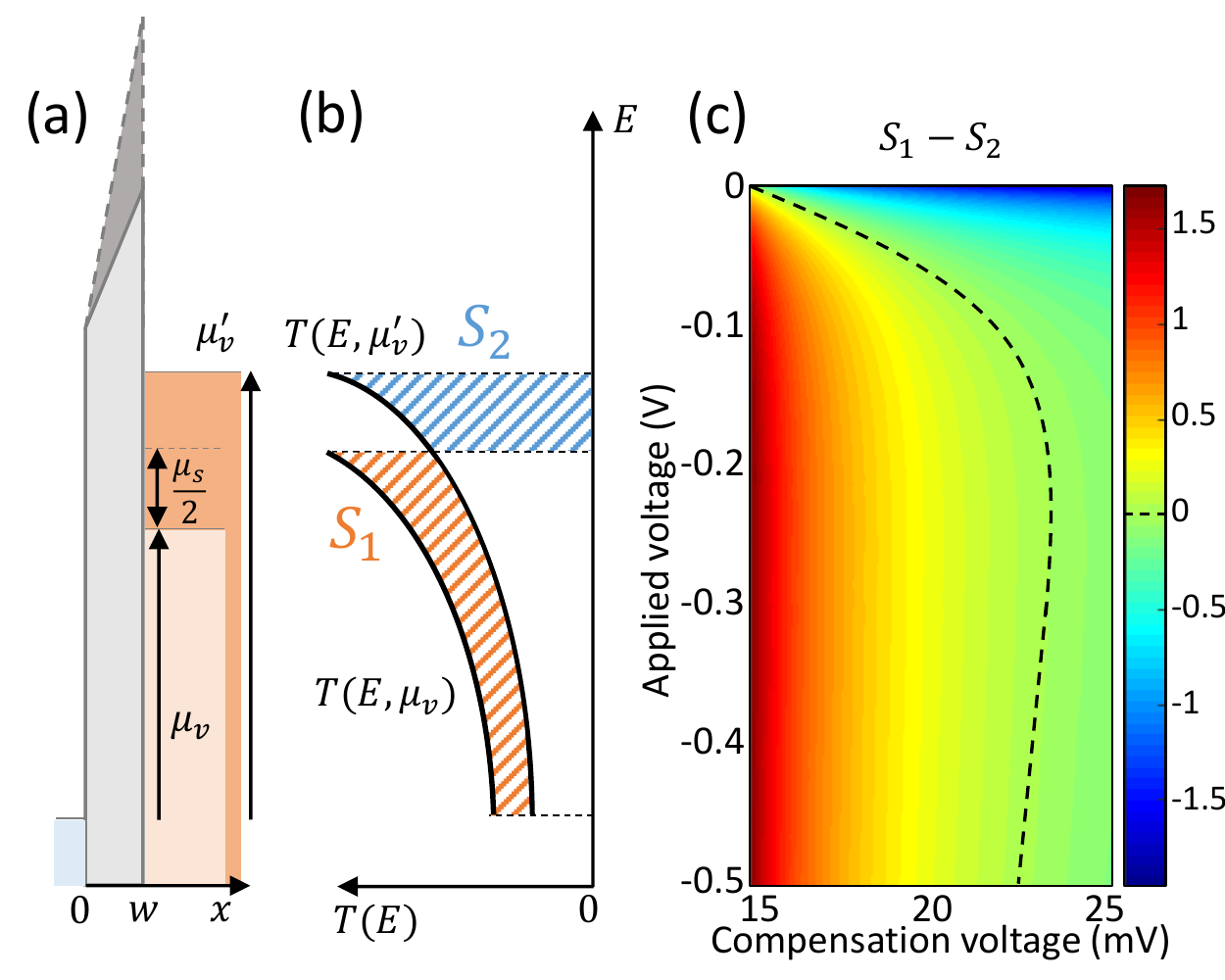}
\caption{(a) Schematic representation of the barrier deformation for two different values of the applied voltage associated to $\mu_v=-eV$ (light color) and $\mu_v^\prime=-eV^\prime$ (dark color). (b) Sketches of the corresponding transmission function under these conditions: the hatched surfaces correspond to the two contribution to current variation due to the bias change. $S_1$ is integrated from $0$ to $\mu_v + \mu_s/2$ and represents the loss of current due to the reduction of the barrier permeability. In the linear regime of spin detection, $S_1=0$. $S_2$ is integrated from $\mu_v+\mu_s/2$ to $\mu_v^\prime$ and corresponds to the gain of current resulting from the enhancement of $V_{\textrm{\tiny spin}}$ in comparison with the linear model. (c) The variation of $S_1-S_2$ with the spin voltage and the applied bias. Along the dashed line, both surfaces are equals and $V_{\textrm{\tiny comp}}=V_{\textrm{\tiny spin}}$.}
\label{Current_Compensation}
\end{figure}

Although the steepness of the transmission function is correlated with the non-linear behaviour of $P_{\textrm{\tiny det}}(V_{\textrm{\tiny app}})$, it is not directly responsible for the change of spin detection efficiency. In this section, we demonstrate that the enhancement of $P_{\textrm{\tiny det}}$ is actually determined by the deformation of the barrier when the device is under an applied bias. Indeed, in a typical spin detection experiment, the applied voltage is compensated with a voltage $V_{\textrm{\tiny spin}}$ to maintain a constant current through the junction with and without spin accumulation (see Eq. (\ref{eqJspin})). As shown in Fig. \ref{Current_Compensation}(a), a variation of the bias leads to a deformation of the barrier, producing a significant change in the transmission function. For negative biases, the barrier height mean value increases when $V_{\textrm{\tiny app}}$ becomes more negative. In the same vein, the voltage compensation $V_{\textrm{\tiny spin}}$ also triggers a reduction of transmission through the barrier, therefore resulting in a positive reinforcement of this spin potential compensation, and then a higher $P_{\textrm{\tiny det}}$.

In Fig. \ref{Current_Compensation}(b), we show a sketch of the energy dependency of the transmission function through a barrier under two different biases, respectively $\mu_v=-eV$ in presence of a spin accumulation $\mu_s$ and $\mu_v^\prime=-eV^\prime$ in absence of spin accumulation. For the sake of simplicity, we will consider a barrier spin polarization of $P_G=100\%$ (the effect of the barrier spin polarization at zero bias will be discussed later). Moreover, without loss of generality, calculations are done for a positive spin accumulation. As presented in Fig. \ref{modelFMBNM}(b), the general behaviour of $P_{\textrm{\tiny det}}(V)$ remains unchanged and the rationale proposed hereafter to explain the enhancement of $P_{\textrm{\tiny det}}$ is valid irrespective the value of $\mu_s$:
the tunnel current in a spin detection experiment is proportional to the integrated transmission $T(E)$ from $E=0$ to the electrochemical potential associated to the applied bias, at low temperature. The surface areas resulting from the integration, which determine the spin detection efficiency, are highlighted in the drawing of Fig. \ref{Current_Compensation}(b). The first one ($S_1$) corresponds to the decrease of the tunnel current resulting from the reshape of the barrier. The second one ($S_2$) corresponds to the gain of current due to the increase of the bias (from $|\mu_s/(2e)|$ until $V_{\textrm{\tiny comp}})$. As a voltage compensation of $|\mu_s/(2e)|$ corresponds to the prediction of the linear model, $S_2$ directly reflects the non-linearity.
The increase of voltage $V_{\textrm{\tiny comp}}$ needed to obtain a perfect compensation of $S_1$ by $S_2$ is the spin voltage $V_{\textrm{\tiny spin}}=(\mu_v -\mu_v^\prime)/(-e)$ that is measured in a spin detection experiment. As soon as $S_2>0$, the response becomes non-linear and the spin voltage will become larger than $P_G\mu_s/2$.

\begin{figure}[h!]
\includegraphics[width=0.49\textwidth]{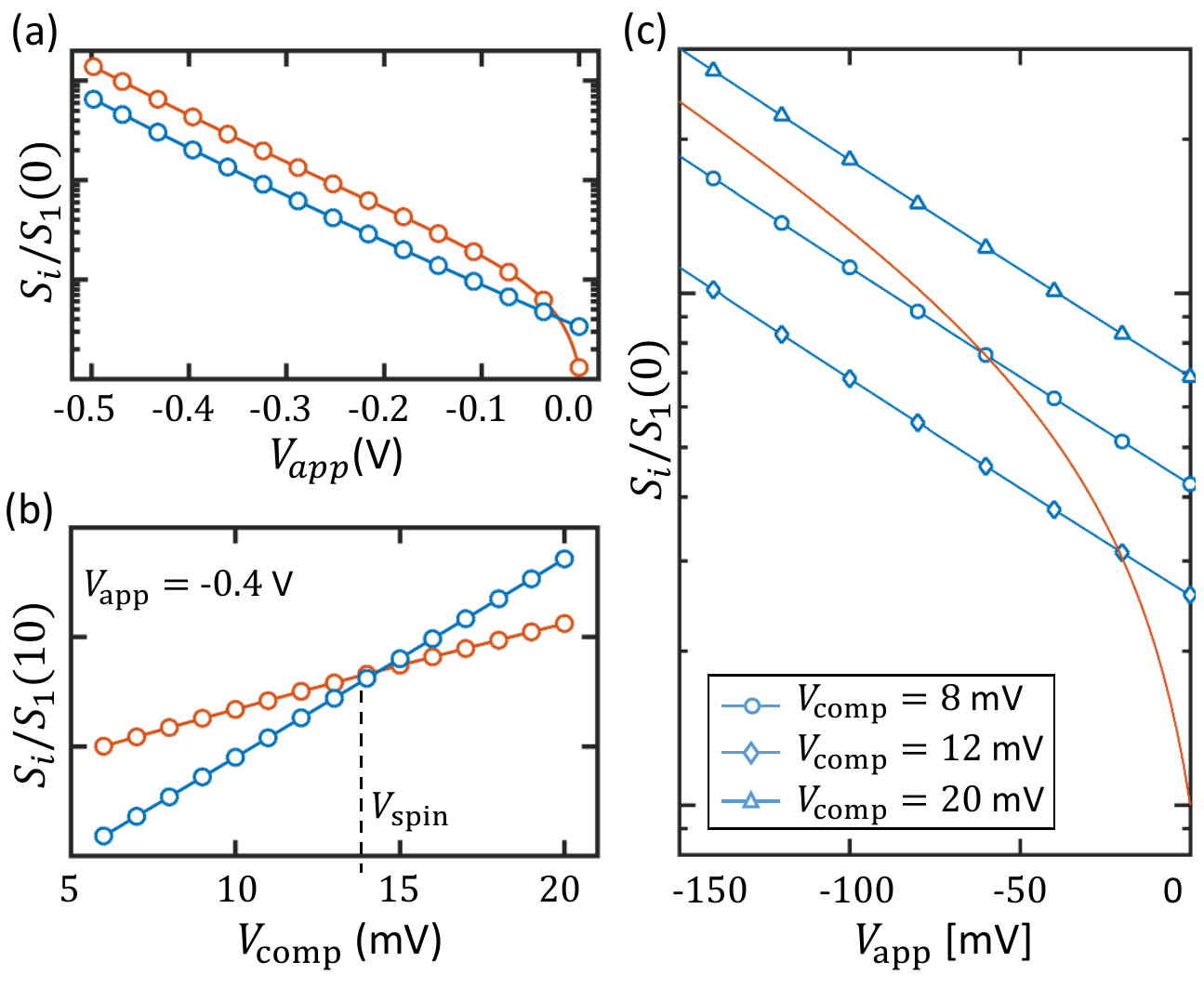}
\caption{Competition between gain and loss of tunnel current in a spin detection experiment bearing non-linear transport. Evolution of integration surfaces $S_1$ and $S_2$ with (a) the applied bias and (b) the compensation voltage. (c) Relative shift of $S_2$ (blue) in comparison to $S_1$ (orange) for different spin voltages.}
\label{S_variation}
\end{figure}

In Fig. \ref{Current_Compensation}(c) we show the difference $S_1-S_2$ for a specific range of bias and voltage compensation. The dashed line denotes the combinations for which both integrals are equals. This line reproduces the behaviour of the spin detection efficiency as a function of the junction voltage, as shown in Fig. \ref{modelFMBNM}(b,c), with a sharp rise followed by a slow decay of the spin detection efficiency as voltage increases in absolute value. As a key consequence, we realize that the relative evolutions of $S_1$ and $S_2$ ultimately determine $P_{\textrm{\tiny det}}$ through a link between the bias and the energy dependencies of the tunnel barrier transmission. In Figs. \ref{S_variation}(a) and (b), we plot the evolution of both surfaces, as function of $V_{\textrm{\tiny app}}$ and $V_{\textrm{\tiny comp}}$, respectively, offering a graphical resolution of the spin detection experiment. 
Such a plot represents a powerful tool to track the value of quantities like $S_1$ and $S_2$ which are essential to understand the electrical spin detection mechanism in presence of a non-linear transport. 

In what follows, we will focus exclusively on negative values of the applied voltage since for this voltage polarity an increase of the spin detection efficiency is expected. First, we observe in Fig. \ref{S_variation}(a) that, for large polarizations, both integrals increase with a similar negative slope, while the slope of $S_1$ decreases abruptly for weak polarizations. In Fig. \ref{S_variation}(b), it is shown that both surfaces increase with the compensation voltage when the applied voltage is kept constant. As demonstrated in the Supplemental Material \cite{SM}, $S_1$ and $S_2$ may be approximated as following: 
\begin{equation}
S_1\approx-V_{\textrm{\tiny comp}}\int_0^{-eV_{\textrm{\tiny app}}+\mu_s/2} T(E,V_{\textrm{\tiny app}})\frac{df(E,V_{\textrm{\tiny app}})}{dV}dE,
\label{eq:S1}
\end{equation}
\begin{equation}
S_2\approx\left(-eV_{\textrm{\tiny comp}}-\frac{\mu_s}{2}\right)T(-eV_{\textrm{\tiny app}},V_{\textrm{\tiny app}}).
\label{eq:S2}
\end{equation}
with $T(E,V)=\exp\left[f(E,V)\right]$.

Based on these equations and on Fig. \ref{S_variation}(b), we observe that $S_2(V_{\textrm{\tiny comp}})$ shows a linear evolution with a slope determined by the transmission probability of a particle with energy $E=eV_{\textrm{\tiny app}}$, while the surface $S_1$ is less sensitive to a change of compensation voltage. Therefore, as shown in Fig. \ref{S_variation}(c), a variation of the compensation voltage ultimately translates into an offset for $S_2$ relatively to $S_1$. In the voltage range where $S_1$ is rapidly varying with $V_{\textrm{\tiny app}}$, an increase of $V_{\textrm{\tiny comp}}$ will raise $S_2$ with respect to $S_1$, combined with a slight increase of the bias voltage for which the intersection $S_1=S_2$ occurs. On the other hand, for larger bias, a change in $V_{\textrm{\tiny comp}}$ will produce a strong increase of the bias at the intersection $S_1=S_2$, due to the fact that $S_1$ and $S_2$ have nearly the same slope. If we refer to Eqs. (\ref{eq:S1}) and (\ref{eq:S2}), a similar slope in logarithmic scale would be possible only if the transmission function $T(E,V)$ is large enough to dominate the integral in the definition of $S_1$. Such is the case if the upper integration limit is large enough.
Considering the previous explanation, the large increase of spin detection efficiency under a low applied voltage, and the slow decrease at higher bias, are related to the deviation of $S_1$ from $S_2$, (i. e., deviation from the exponential nature of $T(E,V)$).

The foregoing argumentation allows us to explain the physical origin of the existence of an optimal value $V_{\textrm{\tiny max}}$ for the spin detection efficiency, as indicated in Fig. \ref{modelFMBNM}(b). On the one hand, the range of electron energy which participate to the current increases with the bias (from $0$ to $\mu_v=-eV_{\textrm{\tiny app}}$, if we assimilate the Fermi-Dirac distribution to a step-like function, as is the case for sufficiently low temperatures). The current gain caused by an increase of the spin voltage is then due to electrons of the higher energy levels. On the other hand, the reduction of barrier permeability impacts the transmission probability of all energy levels. When the range of concerned energy levels is narrow engouh ($V_{\textrm{\tiny app}}$ weak), the transmission function varies slowly with the energy and, subsequently, the participation of low energy electrons is not negligible. 
Therefore, a higher spin voltage is needed to compensate for the current loss due to the reduction of the tunneling capacity of the electrons for all energy levels. However, as the transmission through the barrier evolves exponentially with the energy, the current due to electrons with energies lower than the semiconductor quasi Fermi level ($\mu_v$ in Fig. \ref{Current_Compensation}(a)) becomes less significant. For a certain negative bias, the current gain resulting from the spin voltage compensation overcomes the loss due to the barrier deformation, leading to a decrease of $P_{\textrm{\tiny det}}$. The competition between both effects leads to a maximum spin detection efficiency at a bias $V_{max}$. This result is in agreement with recent experimental observations \cite{Jansen2018}, suggesting that the present descritpion may shed ligth on how to optimze the spin detection efficiency by tuning the barrier parameters. 

Indeed, the energy dependence of the transmission function as well as the way it varies under bias are essential ingredients needed to understand and master the spin detection efficiency.  As explained previously, $P_{\textrm{\tiny det}}$ is sensitive to the barrier properties: width, height and spin polarity.
\begin{figure}[h!]
\includegraphics[width=0.49\textwidth]{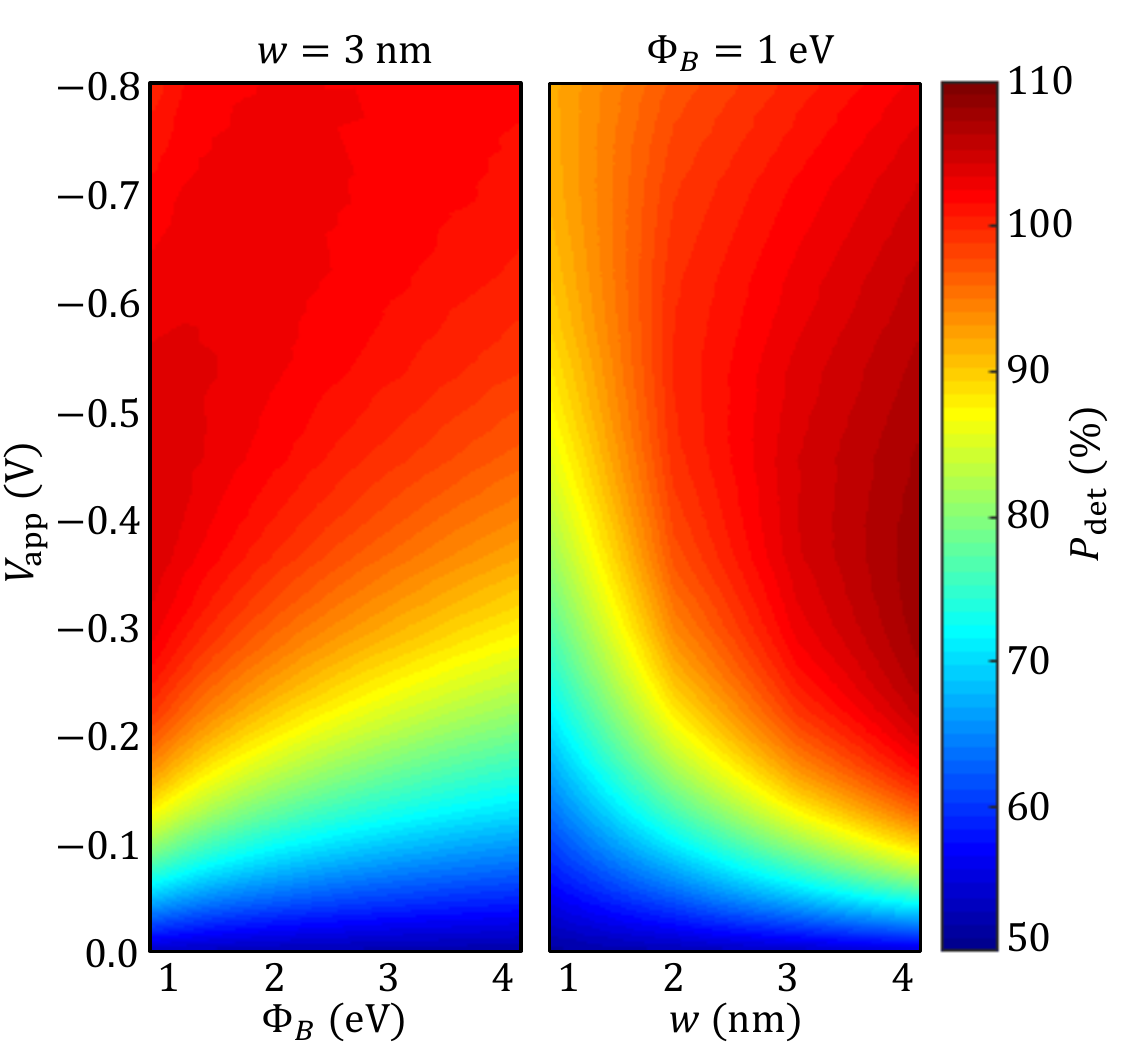}
\caption{Effect of the variation of a rectangular barrier dimensions on the spin detection efficiency. Results are obtained for $P_G=50\%$ and $\mu_s=10$ meV.}
\label{HeightWidthPdet}
\end{figure}
Fig. \ref{HeightWidthPdet} summarizes the dependency of $P_{\textrm{\tiny det}}$ on the barrier dimensions. It can be observed that the variation of the width $w$ and the height $\Phi_B$ impact oppositely $P_{\textrm{\tiny det}}$. 
For weak negative biases, the variation of $P_{\textrm{\tiny det}}$ with $V_{\textrm{\tiny app}}$ is larger for lower and thicker barriers, i. e. for sharper transmission function $T(E)$. It is worth noting that the maximum for $P_{\textrm{\tiny det}}$ follows the same trend, although the corresponding variations are less significant ($10\%$ of deviation around 2 times the barrier spin polarity $P_G =50\%$).

\subsection{Effect of the degeneracy level}

Experimental observations in Ref. \cite{Jansen2018} demonstrated that the spin detection efficiency in a rectangular barrier may overcome the theoretical limit of $2P_G$ to reach spin detection 2.3 times the value predicted by the linear model, and even more when taking into account the drastic reduction of $P_{\textrm{\tiny G}}$ with bias. As the barrier used in that experimental work was rectangular (MgO 2 nm thick \cite{Spiesser2017}), the increase of the spin detection efficiency can not be solely explained by the barrier deformation with the applied bias. Indeed, based on our calculations the spin detection efficiency is expected to be lower than 2 times the barrier spin polarization $P_G$. However, as shown in the following discussion, the observed excess spin detection can be justified by including the effect of the band gap in the model. This mechanism is presented here below for a FM/B/SC structure, where SC is a non-magnetic degenerate semiconductor.
\begin{figure}[h!]
\includegraphics[width=0.49\textwidth]{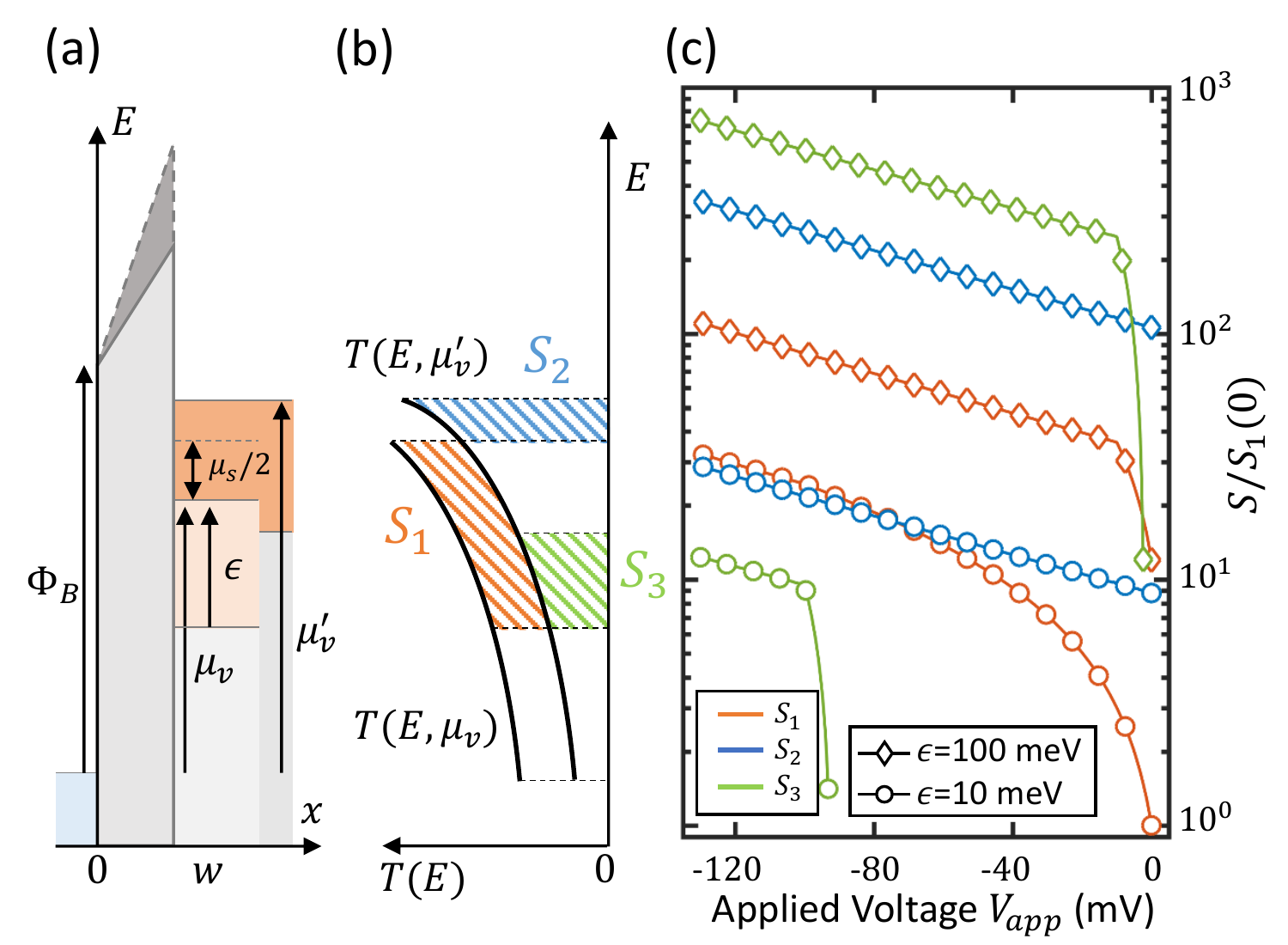}
\caption{Effect of the SC band gap on the spin detection efficiency. (a) Schematic representation of a FM/B/SC structure for a degenerated SC. The Fermi level is located at an energy value of $\epsilon$ above the bottom of the conduction band $E_c$. The barrier and the $E_c$ edge change as the bias is set to $\mu_v^\prime$ (without spin accumulation $\mu_s=0$ - dark color) instead of $\mu_v$ (with spin accumulation - light color). (b) Qualitative sketch of the corresponding transmission function. The $S_3$ integral is associated to the current loss due to the energy shift of the band gap. (c) Quantitative analysis of the variation of the areas $S_1$, $S_2$ and $S_3$ with the applied junction voltage for $V_{\textrm{\tiny spin}}=8$ mV, $\mu_s=10$ meV and for two different values of $\epsilon$. Surfaces (for each $\epsilon$) are normalized by $S_1(0)$ and off-setted.}
\label{bandgap_effect}
\end{figure}

In order to describe the semiconductor with its band gap, we introduce a new parameter, $\epsilon$, that reflects the degree of degeneracy of the semiconductor. $\epsilon$ represents the difference between the electrochemical $\mu_v$ potential and the bottom conduction band. This effect is included in our calculation by introducing an energy dependent DOS $N(E)$ in the NM layer given by
\begin{equation}
N(E)=\frac{8\pi\sqrt{2}}{h^3}m^{3/2}\sqrt{E-(\mu_v+\epsilon)}.
\end{equation} 
The DOS is therefore null for energies in the band gap, i.e. for $E<\mu_v+\epsilon$.
At low temperatures, $\mu_v$ corresponds to the maximal occupied energy level in this band. Therefore, the parameter $\epsilon$ is linked to the carrier concentration in the conduction band by the Fermi-Dirac distribution. Figure \ref{bandgap_effect}(a) sketches the basic case of a sandwich structure FM/B/SC, including the band gap. The consequence of the band gap on the transport of electrons through the barrier is visible when the applied electrochemical potential $\mu_v$ overcomes the degeneracy level $\epsilon$. In this case, electrons that tunnel from the SC into the FM have energies limited by the bottom of the conduction band and not by the occupancy in the FM. As a consequence, when a compensation voltage is applied after removing the spin accumulation, the increase of potential from $\mu_v$ to $\mu_v^\prime$ is accompanied by an increase of the energy level of the bottom of the conduction band. In term of current integration surface, it corresponds to a third surface $S_3$ corresponding to a lost of current due to the shift of $E_c$ induced by a change of applied voltage (see figure \ref{bandgap_effect}(b)). 

\begin{figure*}
\includegraphics[width=0.99\textwidth]{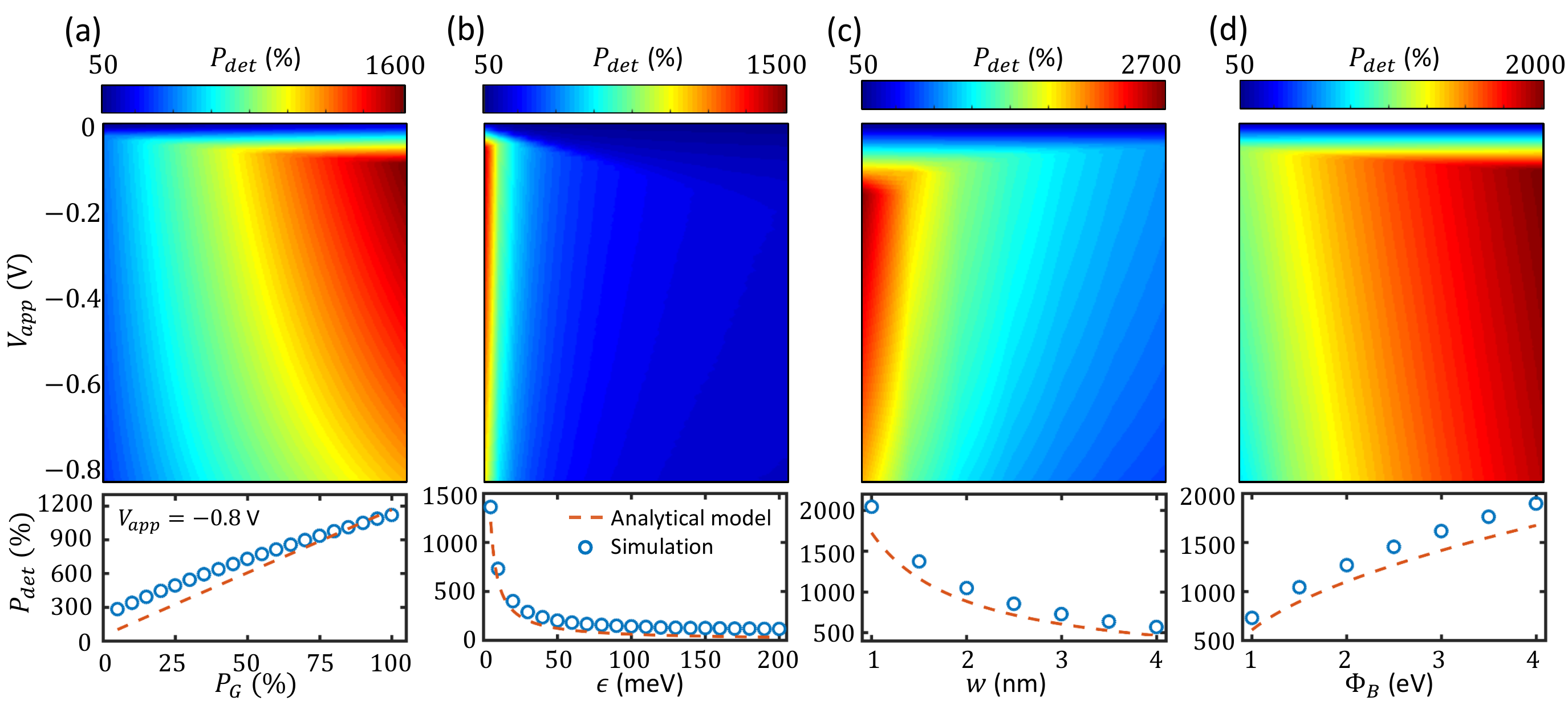}
\caption{Effect of the variation of a rectangular barrier parameters ((a) $P_G$, (b) $\epsilon$, (c) $w$ and (d) $\Phi_B$) on the spin detection efficiency when the band gap affects the tunnel transport. Fixed parameters are $P_G=50\%$, $\epsilon=10$ meV, $w=3$ nm, $\Phi_B=1$ eV. Bottom panels compare results of simulations with the analytical solution from Eq. (\ref{eq:Pdet}) for $V_{\textrm{\tiny app}}=-0.8$ V.}
\label{HeightWidthPdet_bandgap}
\end{figure*}

As demonstrated in the Supplemental Material \cite{SM}, in addition to the new surface $S_3$, the insertion of the band gap will modify the definition of $S_1$
\begin{equation}
\begin{split}
S_1&\simeq -eV_{\textrm{\tiny comp}}\left(\mu_s/2+\epsilon\right)T(-eV_{\textrm{\tiny app}},V_{\textrm{\tiny app}})\frac{df(-eV_{\textrm{\tiny app}},V_{\textrm{\tiny app}})}{dV}\\
S_3&\simeq -eV_{\textrm{\tiny comp}}\left(1-\left(\dfrac{eV_{\textrm{\tiny comp}}}{2}+\epsilon\right)\dfrac{df(-eV_{\textrm{\tiny comp}},V_{\textrm{\tiny comp}})}{dE}\right)\\
&~~\times T(-eV_{\textrm{\tiny app}},V_{\textrm{\tiny app}}).
\end{split}
\end{equation}
Both $S_1$ and $S_3$ contribute to the reduction of the tunnel current and their respective contribution is dependent on the level of degeneracy. As $\epsilon$ decreases, $S_1$ will decreased and $S_3$ will increase (note that $-eV_{\textrm{\tiny comp}}$ is positive and larger than $\mu_s/2$). 
As represented in panel (c), the impact of $S_3$ on the spin detection efficiency is reduced for highly doped SC ($\epsilon$ large). In this case, the increase of the spin detection efficiency is dominated by the barrier deformation. In contrast to that, for SC with a weaker level of degeneracy, the impact $S_3$ may overcome the lost of current due to barrier deformation $S_1$. As a consequence, a spin detection efficiency higher than 2 times the barrier polarity is possible. 

Under the approximation $-eV_{\textrm{\tiny app}}>> \epsilon,\mu_s$, we propose an analytical model to predict the non-linearity of the spin detection efficiency at low temperature. In this case, as demonstrated in the Supplemental Material \cite{SM},
\begin{equation}
P_{\textrm{\tiny det}}\simeq\dfrac{P_G+\dfrac{\mu_s}{4e}\dfrac{df(-eV_{\textrm{\tiny app}},V_{\textrm{\tiny app}})}{d(E/e)}}{\dfrac{\epsilon}{e}\left(\dfrac{df(-eV_{\textrm{\tiny app}},V_{\textrm{\tiny app}})}{d(E/e)}-\dfrac{df(-eV_{\textrm{\tiny app}},V_{\textrm{\tiny app}})}{dV)}\right)}.
\label{eq:Pdet}
\end{equation}
The dependence to the barrier shape and deformation is determined by the partial derivatives of the function $f(E,V)$ evaluated in $E=-eV_{\textrm{\tiny app}}$ and $V=V_{\textrm{\tiny app}}$. As predicted in our simulations, $P_{\textrm{\tiny det}}\sim \epsilon^{-1}$. In the case of n-type silicon substrate, a doping level between $5 \times 10^{18}$ and $1 \times 10^{20}$ cm$^{-3}$ corresponds to an $\epsilon$ between $0.01$ to $0.1$ eV.

In Fig. \ref{HeightWidthPdet_bandgap}, the effect of a variation of the barrier height and width is analysed. The presence of the band gap severely modifies the spin detection response due to the barrier deformation under bias (see figure \ref{HeightWidthPdet}) since now $P_{\textrm{\tiny det}}$ decreases with an increase of width and a decrease of the barrier height. We conclude that the impact of $S_3$ is less important for barriers with a steeper transmission energy dependence. Indeed, for a barrier with a sharp transmission probability, the current due to electrons with a weak energy (range of energy for $S_3$) is negligible in comparison to those of higher energy (range of $S_2$). Therefore a reduction of the energy dependence of the transmission probability acts as an increase of the degeneracy level. 
Results from Fig. \ref{HeightWidthPdet_bandgap} show that $P_{\textrm{\tiny det}}$ reach huge values of several thousands of percent while $P_G$ is only $50\%$. Such a large non-linearity factor may explain the reported deviation of several orders of magnitude between the (linear) theory and the Hanle experiments achieved in 3T devices \cite{Jansen2012}. In the same vein, we assume that this enormous increase of $P_{\textrm{\tiny det}}$ with the reduction of the doping level may also explain why a non-negligeable spin signal is detected in devices where thermally-assisted tunneling occurs \cite{Spiesser2014}. However, it is important to note that a reduction of the doping level may amplify the depletion layer at the interface B/SC, which reduces drastically the detected spin voltage \cite{Jansen2007}. The combination of those effects may explain why a similar spin voltage was obtained in Ref. \cite{Li2011} for different doping levels. 
 
In a tunnel barrier designed for electrical spin detection, if a negative bias is applied, both the deformation of the barrier transmission and the change of energy range for carriers that participate to the transport are responsible for the observation of a colossal non-linear spin detection efficiency. For rectangular barriers, the second phenomenon seems to dominates except for highly degenerate SC. The numerical simulations as well as the analytical approach highlight that the non-linearity of the spin detection is sensitive to the energy dependency of the transmission steepness. Steeper transmission may be obtained if a non-rectangular tunnel barrier is used.
\begin{figure*}
\includegraphics[width=0.99\textwidth]{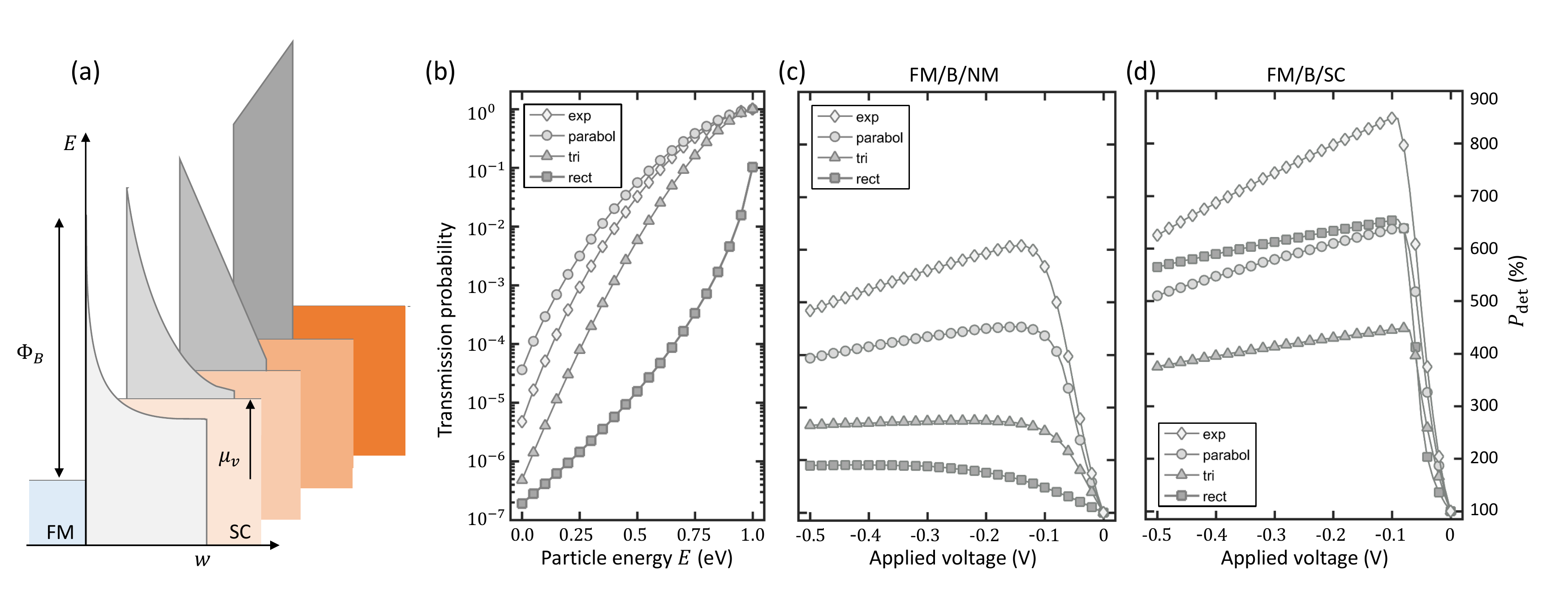}
\caption{(a) Schematic representation of the different shapes of the interface barrier, from the background to the front: rectangular, triangular, parabolic and exponential evolution of the barrier height with the distance from the SC. (b) Energy dependency of the transmission probability through the barrier. (c,d) Non-linearity of the spin detection efficiency with the applied voltage for tunnel barrier with various shapes. Results depicted on panel (c) show the effect of the barrier shape for a FM/B/NM configuration while those shown on panel (d) focus on the case FM/B/SC. Calculations were performed using $w = 3$ nm (excepted for the rectangular barrier where $w=2$ nm); $\Phi_B=1$ eV; $P_G=100\%$; $T=1$ K; $\epsilon=50$ meV.}
\label{Compare_shape}
\end{figure*}
\subsection{Effect of barrier shape}

At low bias, the behaviour of $S_1$ is directly dependent on the barrier shape (see SI). In this section, we would like to quantify the degree of sensitivity of $S_1$ to that critical feature of the interface. The comparison is made for four different shapes, respectively a rectangular, triangular, parabolic and exponential spatial dependence (see Fig. \ref{Compare_shape}(a)). Each barrier is determined by a maximal height $\Phi_B$, a width $w$ and degree of degeneracy $\epsilon$. The way those parameters influence the shape of the barrier is detailed in the SI. The width associated to each barrier (respectively 2, 3, 3 and 3 nm) has been arbitrary chosen in order to have transmission probability in the same range of values. The maximal barrier height at zero current is set to $\Phi_B=1$ eV. 

The transmission function of each barrier (at $V_{\textrm{\tiny app}}=0$) is plotted in figure \ref{Compare_shape}(b). As expected, the steepness of the transmission function for electrons with an energy close to the quasi-fermi level increases from the rectangular barrier to the exponential one.


Spin detection efficiencies under different biases were computed for a constant spin accumulation of $\mu_s=10$ meV (Fig. \ref{Compare_shape}(c,d)). Results are presented for the FM/B/NM and FM/B/SC structures. It allowed us to separate the effect of the barrier shape modification from the effect of the band gap. Indeed, except for the case of the rectangular barrier, a change of $\epsilon$ does lead to a reshaping of the barrier. For the model that does not include the band gap, $P_{\textrm{\tiny det}}$ tends to the value of 2$P_G$ for a rectangular tunnel barrier, while the maximal spin detection efficiency skyrockets when the barrier height depends on the distance from the FM/B interface. 
For the case of a barrier sandwiched between a FM metal and a degenerate SC, the maximal spin detection efficiency is improved irrespective the barrier shape. However the rectangular barrier is more sensitive to a change of $\epsilon$ since it is correlated with the steepness of the transmission function as explained above.  
It is worth noting that a change of $\epsilon$ does modify the spin transport through two different mechanisms. First, it reduces the range of energy of carriers involved in the tunnel transport. Secondly, it changes the shape of the barrier, as it could be expected for a Schottky junction. We conclude that the huge spin detection efficiency improvement due to the non-linearity of the tunnel junction arises in every type of barrier, as previously suggested by Jansen \textit{et al}. \cite{Jansen2018}. However, in contrast to the latter study, we demonstrate that the dominant mechanism varies between an oxide based tunnel junction and a FM/SC contact Schottky junction, which is a nuance that we deem important for understanding the whole picture. 

\subsection{Dependence on $\mu_s$ and spin lifetime}

\begin{figure}[h!]
\includegraphics[width=0.49\textwidth]{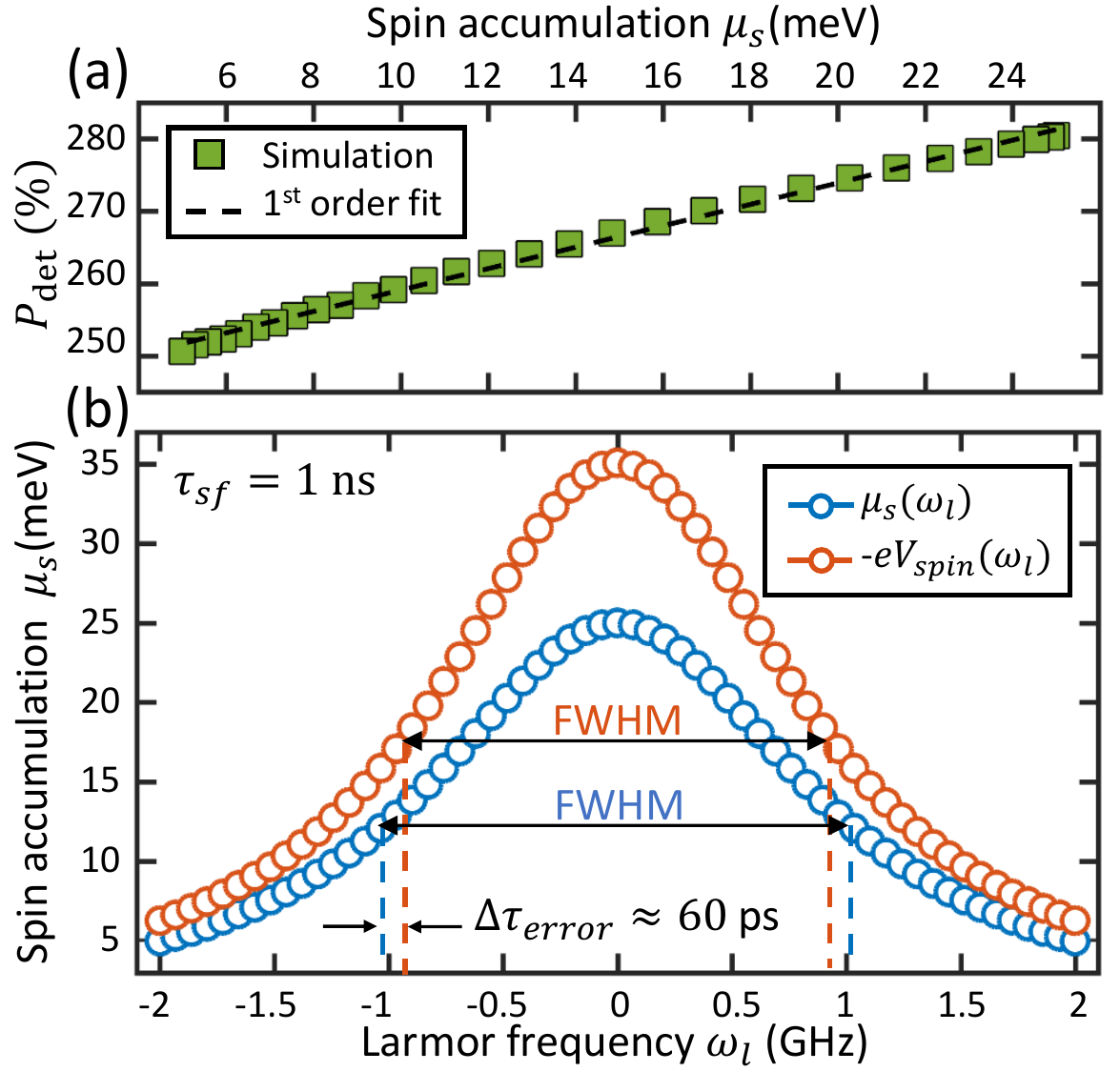}
\caption{Consequences of the variation of the spin detection efficiency with the spin accumulation. (a) Computed value of $P_{\textrm{\tiny det}}$ for different spin accumulations at the vicinity of a tunnel rectangular barrier ($w=3$ nm; $\Phi_B=1$ eV; $P_G=50\%$; $V_{\textrm{\tiny app}}=-200$ mV; $\epsilon=50$ meV). (b) Numerical simulation of the spin precession experiment. $\mu_s$ and $eV_{\textrm{\tiny spin}}$ are plotted with the corresponding spin lifetime assuming that both are Lorentzian distributions.}
\label{HanlePrecession2}
\end{figure}

In addition to the correction of the amplitude of the predicted spin accumulation that is formed in the SC, calculations performed in our study suggest that the predicted spin lifetime in Hanle precession experiments needs to be adjusted. Such a correction arises from the fact that, the applied voltage affects the spin detection efficiency, and therefore the spin accumulation $\mu_s$ will also induce a deviation of $P_{\textrm{\tiny det}}$ from $P_G$. In the non-linear theory, a higher spin voltage is linked to a higher barrier deformation which, in turn, triggers an increase of the spin voltage. Therefore, the spin voltage is expected to deviate from a linear dependence with the spin accumulation. As shown in Fig. \ref{HanlePrecession2}(a), our results highlight this observation and show that $P_{\textrm{\tiny det}}$ is proportional to $\mu_s$. This result obviously impacts the spin diffusion length deduced in Hanle precession measurements. The theory underlying such processes implies that the spin accumulation is destroyed when applying a magnetic field perpendicular to the spin preferential orientation of magnetization \cite{epub1823}. Under a magnetic field $B$, the spin accumulation $\mu_s(B)$ follows a Lorentzian shape with a maximum value $\mu_s(0)$. From the full width at half maximum (FWHM) of the Lorentzian function, one can deduce the spin lifetime of carriers injected into the NM layer, namely $\tau_{sf} = 2/(FWHM)$. However, this kind of experiment is performed on the spin voltage instead of the spin accumulation.

 Therefore, in the non-linear transport regime, a modified Lorentzian distribution is needed. Supposing $P_{\textrm{\tiny det}}=\alpha \mu_s +\beta$ for an applied voltage $V_{\textrm{\tiny app}}$, the variation of $V_{\textrm{\tiny spin}}$ with the magnetic field becomes 
\begin{equation}
V_{\textrm{\tiny spin}}(\omega_L)=\frac{\mu_s(0)}{2\left(1+(\omega_L\tau_{sf})^2\right)} \left[\frac{\alpha\mu_s(0)}{1+(\omega_L\tau_{sf})^2}+\beta\right],
\end{equation} 
where $\omega_L$ is the Landau frequency, linearly dependent on the magnetic field $B$. Consequently, we suggest a correction for the equation that allows one to extract the spin lifetime from Hanle precession measurements, $\tau_{sf}=\frac{2}{FWHM}\sqrt{\frac{\sqrt{\beta^2 +2\alpha\mu_s(0)(\alpha\mu_s(0)+\beta)}-\alpha\mu_s(0)}{\alpha\mu_s(0)+\beta}}$. According to Eq. (\ref{eq:Pdet}), the ratio between $\alpha$ and $\beta$ is
\begin{equation}
\dfrac{\alpha \mu_s(0)}{\beta} = \dfrac{\mu_s}{4eP_G} \dfrac{df(-eV_{\textrm{\tiny app}},V_{\textrm{\tiny app}})}{d(E/e)}
\end{equation}
For rectangular oxide barrier, the correction on the spin lifetime will be limited, around $5\%$ of error. However, in case of tunnel Schottky barrier, the error may increase until $20\%$.

\subsection{Effect of the temperature}
\begin{figure}[h!]
\includegraphics[width=0.49\textwidth]{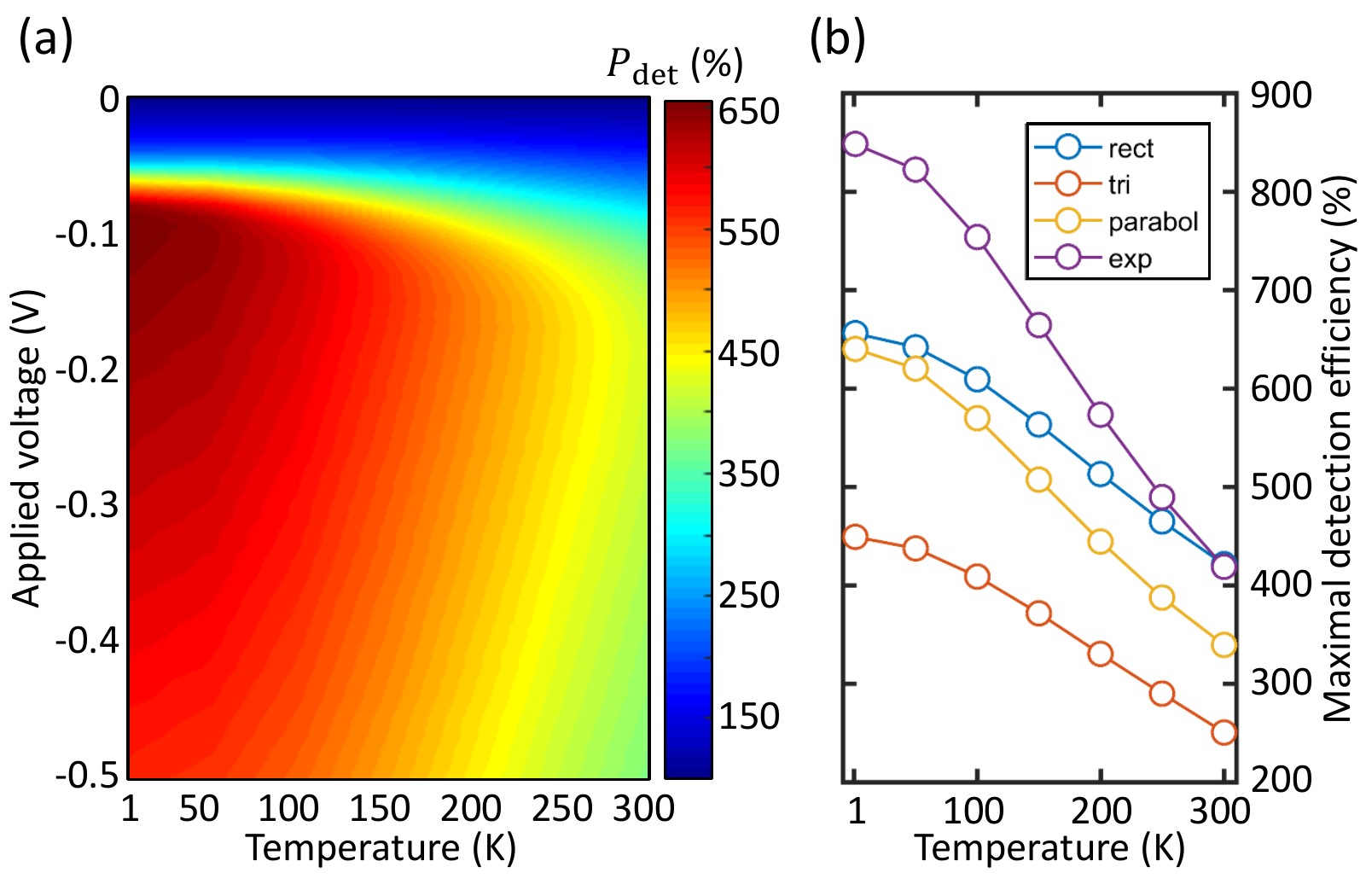}
\caption{Effect of temperature on the spin detection efficiency. (a) Variation of $P_{\textrm{\tiny det}}$ with the temperature and the junction bias. (b) Comparison of temperature effect on different barrier shapes. Calculations were performed using barrier properties as presented in Fig. \ref{Compare_shape}.}
\label{temperature}
\end{figure}

In order to complete our analysis of the non-linearity of the spin detection efficiency under bias, we performed simulations for different temperatures.  In Fig. \ref{temperature}(a),  $P_{\textrm{\tiny det}}(V_{\textrm{\tiny app}})$ is plotted for a range of temperatures from 1 K to 300 K, and shows that the spin detection is less efficient at high temperatures. This behaviour has been systematically observed in 3T devices in which the reduction of the spin voltage with temperature was attributed to an increase of the thermal noise, an increase of the thermionic emission transport and a simultaneous reduction of the spin polarity of the barrier $P_G$ \cite{Spiesser2017}. 
As those effects are not included in the present calculations, the decrease of the detected spin signal with temperature should be ascribed to another phenomenon. More precisely, the temperature dependence of the detection efficiency can be simply linked to the broadening of the Fermi-Dirac function. As the temperature increases, the Fermi-Dirac distribution deviates from the Heaviside step-like function. As a result, carriers with energies slightly higher than the quasi-Fermi level will participate to the charge transport through the tunnel barrier. Consequently, carriers with a higher transmission probability will be involved. Therefore, the gain of current due to the increase of the applied voltage $V_{\textrm{\tiny comp}}$ will be increased. This effect may be seen as an increase of the carrier concentration in the conduction band (i.e., an increase of $\epsilon$), which results in a reduction of the spin detection efficiency.
The results depicted in Fig. \ref{temperature}(b) show the effect of temperature for the different barrier shapes that have been studied in this work. It is noted that the maximal spin detection efficiency decreases whatever the shape of the barrier. At low temperature, the effect is more pronounced is the non-rectangular barrier, suggesting that its origin is related to the steepness of the energy dependence of the transmission function. This outcome tends to confirm the justification that higher temperatures allow to activate carrier with higher energies (associated to higher barrier transmission) and, therefore, that the current is compensated more easily by the increase of the compensation voltage. For barriers with sharp transmission functions, the gain of current due to high energy carriers will obviously be higher.

\section{Conclusion}

 In summary, we demonstrate that the non-linearity of the spin detection efficiency under bias results from two different mechanisms: the tunnel barrier deformation and the conduction band shift, leading to spin detection efficiency higher than 10 times the one predicted by the linear model. As a consequence, we emphasize the necessity to take into account the effect of the energy dependency of the tunnelling transmission probability as well as the band gap (even for highly degenerate SC) in the model used to analyse results from local (2T and 3T) spin devices. Effects of the barrier shape, the doping level and the temperature on the magnitude of the probed spin accumulation and spin relaxation time have been studied, leading to a better interpretation of spin detection experiments. We believe that our results may clarify  the complex mechanisms that govern spin injection, transport and detection experiments and may possibly explain numerous puzzling results reported in the literature.  

\section{Acknowledgements}
Financial support by the ARC grant 13/18-08 for Concerted Research Actions, funded by the Wallonia-Brussels Federation, is gratefully acknowledged.
\bibliography{biblio}
\end{document}